\documentclass[showkeys,showpacs,aps,twocolumn]{revtex4-1}

\usepackage{appendix}
\usepackage{hyperref}
\usepackage{graphicx}
\usepackage{amsmath}
\usepackage{yfonts}
\usepackage{graphicx}
\usepackage{color}
\usepackage{ctable}
\usepackage{grffile}
\usepackage{epstopdf}
\usepackage{subfig}
\usepackage{subfloat}

\def\neff{N_{\rm eff}}

\def\be{\begin{equation}}
\def\ee{\end{equation}}
\def\la{\label}
\def\bea{\begin{eqnarray}}
\def\eea{\end{eqnarray}}

\def\fr{\frac}
\def\le{\left}
\def\ri{\right}

\def\la{\label}

\begin{document}

\title{Extra relativistic degrees of freedom without extra particles using Planck data}
\author{Jorge Mastache}
\author {Axel de la Macorra}
\affiliation{Instituto de Fisica, Universidad Nacional Autonoma de Mexico, Apdo. Postal 20-364, 01000, Mexico, D.F.}
%\email{mastache@fisica.unam.mx}

\begin{abstract}
  A recent number of analysis of cosmological data have shown indications for the presence of extra radiation beyond the standard model at equality and nucleosynthesis epoch, which has been usually interpreted as an effective number of neutrinos, $\neff > 3.046$. In this work we establish the theoretical basis for a particle physics-motivated model (Bound Dark Matter, BDM) which explain the need of extra radiation. The BDM model describes dark matter particles which are relativistic at a scale below $a < a_c$, these particles acquire mass with an initial velocity, $v_c$, at scales $a > a_c$ due to non-perturbative methods, as protons and neutrons do, this process is described by a time dependent equation of state, $\omega_{\rm bdm}(a)$. Owing to this behavior the amount of extra radiation change as a function of the scale factor, this entail that the extra relativistic degrees of freedom $N_{\rm ex}$ may also vary as a function of the scale factor. This is favored by data at CMB and BBN epochs. We compute the range of values of the BDM model parameters, $x_c = a_c v_c$, that explain the values obtained for the $^4$He at BBN and $\neff$ at equality. Combining different analysis we compute the value $x_c = 4.13(^{+3.65}_{-4.13}) \times 10^{-5}$ and $v_c = 0.37^{+0.18}_{-0.17}$. We conclude that we can account for the apparent extra neutrino degrees of freedom $N_{\rm ex}$ using a phase transition in the dark matter with a time dependent equation of state with no need for introducing extra relativistic particles.
\end{abstract}

\maketitle

\section{Introduction}\label{sec:introduction}
% CMB
In recent years, precision measurements have reveal an incredible amount of information to describe the Universe, inter alia, the fluctuations in the cosmic microwave background (CMB) radiation \cite{Ade:2013lta,Hinshaw:2012fq}, the type Ia supernova \cite{Riess:2011yx}, the large-scale structure (LSS), and the baryon acoustic oscillations (BAO)\cite{Percival:2009xn}.

Cosmological observations are systematically consistent with the standard model of cold dark matter (CDM) with a cosmological constant driving an accelerated expansion of the universe. Nevertheless, there are few unresolved issues, aside from the nature of dark matter (DM) and energy, that could be evidence of physics beyond the  paradigm of CDM, some of them are the large amount of substructure and the incompatible cuspy energy density profile for the galaxy DM halo predicted by CDM framework. Another problem is the amount of radiation in the Universe, therefore, the expansion rate at the early Universe.

Among the fundamental observables that describe the universe lies the redshift of equivalence, $z_{eq}$, the moment when the energy density of matter and radiation were equal. This equality epoch is relevant in structure formation and it also affects the early Integrated Sachs-Wolfe (ISW) effect constrains the value of the matter-radiation equality. The later the equality epoch is, the more the ISW effect CMB photons receive. The early ISW effect is actually a directly constraint via the ratio from the first and third peak of the CMB spectrum \cite{Komatsu:2008hk}, therefore we can compute the amount of radiation at equality if we know the amount of matter in the universe and viceversa.

The radiation energy density at equality epoch is given by photons and neutrinos. The standard model neutrino species contributes $N_{\rm \nu} = 3.046$ degrees of freedom \cite{Mangano:2005cc}. However recent analysis of the Planck \cite{Ade:2013lta} data, WMAP \cite{Hinshaw:2012fq}, Atacama Cosmology Telescope \cite{Sievers:2013wk}, the South Pole Telescope \cite{Keisler:2011aw}, the Sloan Digital Sky Survey (SDSS) data release 7 (DR7) \cite{Hamann:2010pw,Reid:2009nq} and several other analysis \cite{Hamann:2011hu,Archidiacono:2011gq} have reported indication that the effective degrees of freedom, $\neff \equiv N_{\rm \nu} + N_{ex}$, could be greater than the expected, $\neff > N_{\rm \nu}$. The value of $\neff$ obtained before Planck data had an $N_{ex}>0 $ at $1\sigma$ and consistent with $N_{ex}=0 $ at $2\sigma=0$ however the new Planck data has an $N_{ex}=0 $ at $1\sigma$  but the central value hints for a small amount of extra relativistic degrees of freedom. It is worth noticing  that there is a tension between the value of the Hubble constant $H_o$ (in units of ${\rm km\, s^{-1}\, Mpc^{-1}}$) with $H_o = 67.3 \pm 1.2$ for Planck \cite{Ade:2013lta},  $H_o = 70.0 \pm 2.2$ for WMAP9 \cite{Hinshaw:2012fq} and $H_o= 74.8 \pm 3.1$ for Cepheids+SNeIa \cite{Riess:2011yx}, and the value of $\neff$ depends strongly on it. The larger $H_o$ the more amount of relativistic degrees of freedom is required. In order to have a wider scope we present our results using different set of data. This means that the amount of radiation prior the epoch of decoupling seems to be more than the expected ($\hat{\rho}_r > \rho_r$), where $\hat{\rho}_r  = \rho_r + \rho_{\rm ex}$ and $\rho_{\rm ex}$ identifies an extra relativistic component.

Additionally, recent studies find a somewhat higher $^4$He abundance of $Y_p > 0.25$ \cite{Hinshaw:2012fq,Izotov:2010ca,Aver:2010wq} also suggesting a novel radiation during the Big Bang Nucleosynthesis (BBN) epoch. In both cases the extra relativistic component is parameterized as extra neutrino degrees of freedom as function of the neutrino temperature. The value of $\neff$ obtained from data analysis gives a larger value at BBN  than that at equality. In a CDM scenario,  extra relativistic particles give a constant contribution to  $\neff$.

Constraints on $\neff > N_\nu$ can be interpreted as the existence of radiation energy beyond the standard model. In the literature there are different proposals in order to explain the extra radiation, some of them propose the existence of new particles, for instance sterile neutrinos \cite{Hamann:2010bk,deHolanda:2010am}, or proposing that the radiation is a relativistic product of a massive relic particle \cite{Ichikawa:2007jv,Hasenkamp:2011em,Graf:2013xpe}. We also know that $\neff$ cannot be accounted by statistical effect alone \cite{Hamann:2011hu}.

Here we are going to propose a different interpretation of the extra radiation, we shall see that a model named Bound Dark Matter (BDM) \cite{delaMacorra:2009yb}, where the DM particles go through a phase transition,  can well explain the need for an extra component without introducing new particles. The BDM model consists in DM massless particles above a threshold energy ($E_c$) and these particles acquire mass at $E_c$ due to non-perturbative methods, as protons and neutrons do, and their mass is related to its binding energy. We expect this transition to be between an energy of MeV and eV scales, i.e. between  BBN and matter-radiation equality epoch. If $E_c > MeV$ then  BDM would reduce to CDM and for $ E_c < E_{\rm eq}=\mathcal{O}(eV) $ there would be no DM to account for structure formation.

\begin{figure}[t]
  \includegraphics[width=0.45\textwidth]{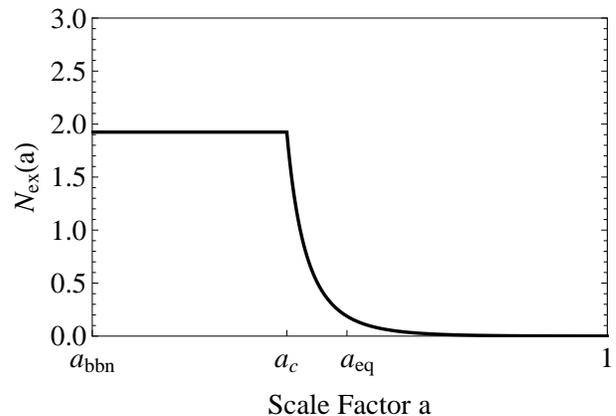}
  \caption{\footnotesize{Plot of the extra relativistic degrees of freedom, $N_{\rm ex}(a)$, as a function of the scale factor using data from Planck. The BDM particles goes through a transition at the $a = a_c$ expected to be smaller than the equality $a_c < a_{\rm eq}$. Before this moment the BDM particles behaves as pure radiation. After the transition the amount of radiation given by the BDM particles is given by Eq.(\ref{eq:rho_ex}). This means that the extra degrees of freedom remains constant before the transition, and after that decrease as a function of the scale factor, c.f. Eq.(\ref{eq:nex}).}}
  \label{fig:Nexfdea}
\end{figure}

We expect that the BDM model describes relativistic particles at BBN epoch and below $E_c$ this particles go through a phase transition between radiation and matter described by a time dependent equation of state (EoS) and this particular behavior allow us to understand the inconsistency in the number of degrees of freedom of neutrinos at equality and BBN epochs by predicting that the amount of extra radiation change as function of the scale factor, if this extra radiation is written in terms of the neutrino radiation this means that the neutrino relativistic degrees of freedom will also be a function of the scale factor.
However, since we expect that the BDM phase transition takes place at energy scales $E_c\gg E_{eq}$ the amount of extra degrees of freedom for BDM is small in accordance with recent Planck data, see Fig.\ref{fig:Nexfdea}. We also expect that the transition of the BDM particles influence the CMB power spectrum by changing the time when matter-radiation equality holds. Furthermore, BDM  can reduce the amount of substructure predicted by CDM  \cite{delaMacorra:2009yb} and have a DM with a core galaxy profile \cite{Mastache:2011cn,delaMacorra:2009yb,delaMacorra:2011df}, thus avoiding the CDM problems.

In this paper we present in Sec. \ref{sec:framework} the theoretical framework to compute the matter-radiation equality for the standard CDM and BDM scenario. Current observational evidence also suggests extra radiation at the BBN epoch, we explore the possibility that the BDM particles also account for such an excess in Sec. \ref{sec:BBN}. We compute the range of values of the free parameters of the BDM model using published values of $\neff$ in Sec. \ref{sec:analysis}. Finally, our conclusions are discussed in Sec. \ref{sec:Discussion}.

\section{Framework}\label{sec:framework}
\captionsetup{margin=5pt,font=small,labelfont=bf,justification=centerlast,indention=.25cm}

The growth of structure formation and cosmological observational data show that the scale factor at matter-radiation equality may be different (larger) than the one of the standard CDM scenario with no extra particles. This can be achieved having extra relativistic particles in a CDM model or by having a time dependent EoS for DM as in our BDM model. We will present here a novel and simple way to compute the epoch of  matter-radiation equality, also valid when the EoS of DM  is time dependent, as is the case for the BDM model.

% Standard Scenario, CDM
\subsubsection{CDM Scenario}
First let us present the standard CDM scenario, where DM is given by a massive particle with vanishing dispersion velocity at scales relevant for structure formation. In this scenario we have a non-relativistic energy density  given by the baryons(\emph{b}) and CDM, $\rho_m=\rho_b+\rho_{cdm}$, and the relativistic particles at energies below neutrino decoupling are the photons ($\gamma$) and neutrinos ($\nu$) with an energy density given by
\begin{equation}\label{eq:rho_r}
 \rho_{r} = (1 + \alpha \neff)\rho_\gamma
\end{equation}
where we use the relation $T_\nu = (4/11)^{1/3}T_\gamma$ derived from the entropy conservation across the electron-positron annihilation and $\alpha \equiv  (7/8) (4/11)^{4/3}\simeq 0.227$. If we have extra degrees of freedom  it is common  to parameterized them using  $\neff \equiv N_{\rm \nu} + N_{\rm ex}$, where $N_{\rm \nu}=3.046$ is the neutrinos degrees of freedom and $N_{\rm ex}$ accounts the amount of extra radiation. In the CDM framework considering no extra radiation the equality epoch is ($a_{eq}/a_o =\rho_{ro}/\rho_{mo}$). It is clear that if we have more relativistic degrees of freedom the equality epoch $\hat{a}_{\rm eq}$ (with $\hat{a}_{\rm eq}/a_o =\hat\rho_{ro}/\rho_{mo}$),  will change to
\begin{equation}\label{eq:aeq3}
  \fr{\hat{a}_{\rm eq}}{a_{\rm eq}}= \frac{\hat{\rho}_{ro}}{\rho_{ro}} = \fr{1+ \alpha N_{\rm eff}}{1+ \alpha N_{\rm \nu}}=  1 + \fr{\alpha N_{\rm ex}}{1+ \alpha N_{\rm \nu}},
\end{equation}
where $\hat{\rho}_r = \rho_r + \alpha N_{\rm ex} \rho_\gamma$ is the relativistic energy density including the extra radiation. Clearly, if the measurements give a larger scale factor at equality,  $\hat{a}_{\rm eq}\geq a_{\rm eq}$,  we would then have extra degrees of freedom, $N_{\rm ex} \geq  0$ ($\neff \geq N_{\rm \nu}$) and the equality holds for $N_{\rm ex} =0$.

Before presenting the second scenario where DM is given by BDM instead of CDM let us determine the value of the full EoS given by the total energy density ($\rho_{\rm tot}$) and pressure ($P_{\rm tot}$), $\omega_{\rm tot} \equiv P_{\rm tot}/\rho_{\rm tot}$. For simplicity we will assume that the contribution of Dark Energy (DE) at matter-radiation equality is negligible and we do not expect it to play a significant role in our analysis, of course different DE models, such as early DE, could also be studied. The equation of state, $\omega_{\rm tot}$, of a fluid consisting of relativistic and matter (cold) particles as function of the scale factor is
\begin{eqnarray}\label{eq:omega_total}
  \omega_{\rm tot} = &&\fr{\rho_r/3}{ \rho_r +\rho_m} = \frac{1}{3}\frac{a_{\rm eq}}{{a_{\rm eq}+a}},  \\
    \hat\omega_{\rm tot} =&& \fr{\hat\rho_r/3}{ \hat\rho_r +\rho_m}= \frac{1}{3}\frac{\hat{a}_{\rm eq}}{{\hat{a}_{\rm eq}+a}}.
\end{eqnarray}
Notice that $\omega_{\rm tot} = 1/6$ at equality $a=a_{\rm eq}$ and $\hat\omega_{\rm tot} = 1/6$ at $a=\hat{a}_{\rm eq}$. We propose to use this alternative criterion to define the equality between matter and radiation epoch which is very useful  when particles do not  have a constant EoS at this time. This is the case for  our BDM model where the particles
are in transition between $\omega_{\rm bdm}=1/3$ and $\omega_{\rm bdm}=0$.

% BDM scenario
\subsubsection{BDM Scenario}
\captionsetup{margin=5pt,font=small,labelfont=bf,justification=centerlast,indention=.25cm}
\begin{figure}[t]
  \includegraphics[width=0.5\textwidth]{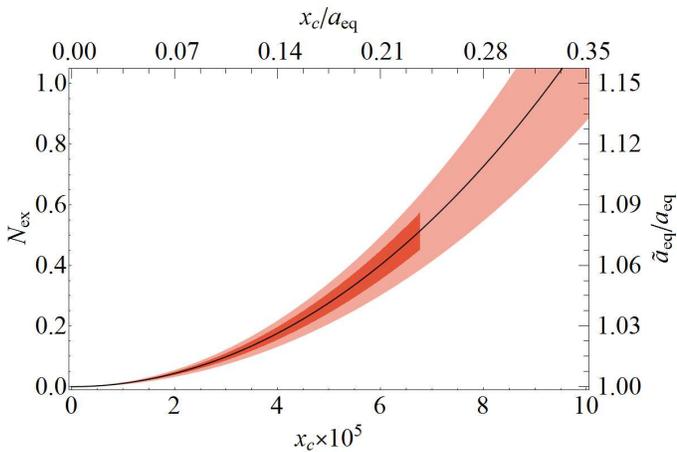}
  \caption{\footnotesize{Plot of $N_{\rm ex}$ and $\tilde{a}_{\rm eq}/a_{\rm eq}$ as a function of $x_c$, see Eq.(\ref{eq:aeqbdm}) and (\ref{eq:nex}) using Planck data, c.f.\ref{tab:results}. The colored region represent two-dimensional (68\%, 95\%) contours marginalized over $\Omega_{mo}$. The thick line represents the central value obtained with the data. We expect $a_c \ll a_{eq}$ and therefore our BDM would only be able to account for a small amount of $N_{ex}$.}}
  \label{fig:omegabdm}
\end{figure}

Let us now present the second scenario where we use our BDM model as DM. The particles of the BDM model go through a non-perturbative phase transition at $a = a_c$ when $\rho_{\rm bdm}(a_c) = \rho_{\rm c} \equiv E_c^4$. At this time the particles acquire mass through non-perturbative phenomena, as protons and neutrons do in QCD. Below the scale $a<a_c$ (or $\rho_{\rm bdm}> \rho_{\rm c}$) they are relativistic ($\omega_{\rm bdm} = 1/3$) massless particles. Above the scale $a>a_c$ the EoS of BDM is time dependent and goes from the values $\omega_{\rm bdm}(a_c) \leq 1/3$ to $\omega_{\rm bdm} \simeq 0$ for $a\gg a_c$.

For simplicity we will take these particles to have an average momentum $\langle|\bar p|^2\rangle$  and average energy  $\langle E\rangle$ so that the pressure becomes $P=n \langle |\bar p|^2 \rangle/3\langle E \rangle$
and the energy density $\rho= \langle E\rangle\, n$, with $n$ the particle number density.   The  EoS for BDM becomes then
\begin{equation}\label{wbdm}
\omega_{\rm bdm} = \frac{ \langle |\bar p|^2\rangle}{3\langle E\rangle^2}= \fr{v_{\rm bdm}^2}{3}= \frac{1}{3} \left(\frac{v_c a_c}{a}\right)^2
\end{equation}
where $v_{\rm bdm}$ is the average velocity of the BDM particles, and we have taken into account that
in a Friedmann-Robertson-Walker background the velocity redshifts with the scale factor as
\begin{equation}\label{vbdm}
v_{bdm}(a)=v_c\left(\frac{a_c}{a}\right).
\end{equation}
The last equation contains two free parameters, the acale factor at the transition $a_c$ and the velocity of the dark particle at that moment, $v_c$. The quantity $v_c$, with $0\leq v_c \leq 1$, gives the initial speed of the particles after the BDM phase transition reflecting the fact that  the BDM particle mass has a non perturbative origin and the resulting velocity may be suppressed in comparison with the speed of light $v_{bdm}(a_c) = v_c < 1$. This is one of the main differences between BDM particles and a standard relativistic particle with a (perturbative) mass $m$ which becomes non-relativistic at $\omega=T/m$ (e.g. at $a=a_c$) with $v=1$. Clearly our BDM reduces to standard CDM when $a_c\rightarrow 0$ (with $v\rightarrow 0$) therefore $\omega_{\rm bdm}\rightarrow 0$, and BDM particles become cold for $a \gg a_c $ and again we have $v\rightarrow 0$ and $\omega_{\rm bdm}\rightarrow 0$. On the other hand, BDM  reduces to a standard particles becoming nonrelativistic at $a_c$ if  $v_c=1$.

Using eq.(\ref{wbdm}) we can integrate $\dot\rho=-3H(\rho+P)$ to obtain an analytic form for $\rho_{\rm bdm}(a)$ which describe the transition of a radiative fluid becoming a matter-like particles. Thus, it allows one to compute the evolution of the expansion rate and cosmological distances easily,
\begin{eqnarray}\label{eq:rho_bdm}
  \begin{alignedat}{2}
    \rho_{\rm bdm} =  &\rho_{\rm c} \left( \frac{a}{a_c} \right)^{-4},\,\, \,\, w=1/3 & \,\, {\rm for} \,\, a < a_c\\
    \rho_{\rm bdm} =  &\rho_{\rm cdm} f(a) & \,\, {\rm for} \,\, a \geq a_c  \\
    f(a)\equiv  & \exp\left[{\frac{3}{2} \omega_{\rm bdm}(a)\left( 1 - \frac{a^2}{a_o^2}\right)}\right]&
  \end{alignedat}
\end{eqnarray}
with $\rho_{\rm cdm}\equiv \rho_{\rm cdmo}\le(a/a_o\ri)^{-3}$, and $\rho_{\rm bdmo}=\rho_{\rm cdmo}=\rho_{\rm dmo}$ is the amount for DM density today, and
\begin{equation}\label{rc}
   \rho_c \equiv E^4_c \simeq  \rho_{\rm bdm o} \left( \frac{a_c}{a_o} \right)^{-3}e^{v_c^2/2}
\end{equation}
where we have used that $f(a_c)\equiv f|_{a=a_c}\simeq {\rm exp}[v_c^2/2]$,  since  $\omega_{\rm bdm}(a_c)=v_c^2/3$ and we have taken
$a_c\ll a_o$, and $f(a_c)$  is now only a function of $v_c$.

\captionsetup{margin=5pt,font=small,labelfont=bf,justification=centerlast,indention=.25cm}
\begin{figure}[b]
  \includegraphics[width=0.50\textwidth]{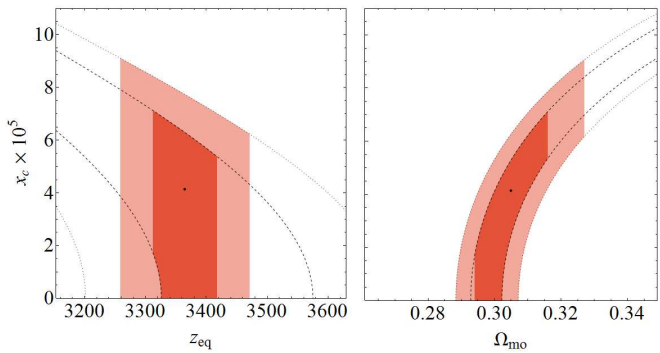}
  \caption{\footnotesize{ In the left panel we show the marginalized two-dimensional (68\%, 95\%) contour in the $x_c$ - $z_{\rm eq}$ plane using Planck results, c.f. Table\ref{tab:results}. The C.L. are trimmed with the respectively $1\sigma$ and $2\sigma$ errors of $z_{\rm eq}$. The dependance on $x_c$ is due to the 1$\sigma$ (dashed) and 2$\sigma$ (dotted) on $\Omega_m$. The point is the central value. In the right panel a similar two-dimensional contour is seen for the $x_c$ - $\Omega_{mo}$ plane where the C.L. are trimmed with the respectively $1\sigma$ and $2\sigma$ errors of $\Omega_{mo}$, and the dependance on $x_c$ is due to the 1$\sigma$ (dashed) and 2$\sigma$ (dotted) on $z_{\rm eq}$.}}
  \label{fig:CL}
 \end{figure}

   \setlength{\tabcolsep}{0.4cm}
   \def\arraystretch{1.25}
   \ctable[
     caption={We present previous results of different surveys when they considered the effective degrees of freedom of the neutrino $\neff$ and/or the primordial helium $Y_p$ as free parameter. The dash means that they consider the fixed value of $Y_p = 0.24$ and/or $\neff = 3.046$. We also show the derived value $\neff^{\rm bbn}$ assuming that the extra radiation was in thermal equilibrium with the photons.},
    label=tab:data,
    star,
    width=0.8\textwidth,
    captionskip=\smallskipamount]{lcccc}
    {  \tnote[a]{This results considered Planck+WMAP9 Polarization(WP)+high-$l$ Planck temperature (highL)+BAO combined data. \cite{Ade:2013lta}.}
       \tnote[b]{This results considered Planck+WP+highL+BAO+HST combined data. \cite{Ade:2013lta}.}
       \tnote[c]{This results considered WMAP9+ACT+SPT+BAO+HST combined data. \cite{Hinshaw:2012fq}.}
       \tnote[d]{Takes into account ACT+WMAP7+SPT+BAO+HST \cite{Sievers:2013wk}.}
       \tnote[e]{Combined data of SPT+WMAP7+BAO+HST \cite{Keisler:2011aw}.} }
    {
                                                  & $\Omega_m$   &  $z_{eq}$     & $\neff$              & $Y_p$           \\
      \hline
      \hline
 \footnotesize{Planck}\tmark[a]+ . . . . . . . .   & $0.308 \pm 0.010$  & $3366 \pm 39$ & --              & --           \\
 \footnotesize{Planck}\tmark[a]+ $\neff$ . . . . . & $0.304 \pm 0.011$  & $3354 \pm 42$ & $3.30 \pm 0.27$ & --           \\
 \footnotesize{Planck}\tmark[a]+ $Y_p$ . . . . . . & $0.306 \pm 0.011$  & $3373 \pm 40$   & --            & $0.267 \pm 0.020$  \\
 \footnotesize{${\rm Planck}_\star^a$}+ $\neff$ + $Y_p$ . . & $0.305 \pm 0.011$  & $3365 \pm 53$ & $3.19_{-0.43}^{+0.54}$ & $0.260^{+0.034}_{-0.029}$   \\
 \footnotesize{Planck}\tmark[b]+ $\neff$ . . . . . & $0.296 \pm 0.010$  & $3329 \pm 38$ & $3.52 \pm 0.24$ & --            \\
 \footnotesize{WMAP9}\tmark[c] . . . . . . . .     & $0.287^{+0.009}_{-0.009}$   & $3318 \pm 55$ & $3.55_{-0.48}^{+0.49}$ & $0.278^{+0.034}_{-0.032}$   \\
 \footnotesize{ACT}\tmark[d]. . . . . . . . . . .  & $0.29 \pm 0.01$  & $3312 \pm 78$ & $3.50 \pm 0.42$ & $0.255^{+0.01}_{-0.11}$ \\
 \footnotesize{SPT}\tmark[e]. . . . . . . . . . .  & $0.28 \pm 0.02$  & $3267 \pm 81$  & $3.86 \pm 0.42$ & $0.296^{+0.30}_{-0.30}$ \\
      \hline
  }

% Omega = 1/6 method
It is naive to think that the matter-radiation ratio at early ages can be computed by a simple extrapolation of today's values because we are proposing a DM phase transition with a time dependent $\omega_{\rm bdm}$. Hence, we cannot say that matter-radiation equality is when $\rho_r = \rho_{m}$ but instead we define equality  when the total EoS is $\omega_{\rm tot} = 1/6 $, as discuss in Eq.(\ref{eq:omega_total}), that overcomes the fact that $\omega_{\rm bdm}$ is a function of $a$. Therefore in the case of the BDM we have,
\begin{eqnarray}
      \omega_{\rm tot}(a) &=& \fr{\rho_r/3+\omega_{\rm bdm}\rho_{\rm bdm}}{\rho_r +\rho_b+\rho_{\rm bdm}}   \nonumber\\
                       &=& \fr{1/3+\fr{a}{a_{eq}}\fr{\Omega_{bdm o}}{\Omega_{mo}} \omega_{\rm bdm}(a) f(a)}{ 1+\fr{a}{a_{eq}}\le(1 + \fr{\Omega_{bdm o}}{\Omega_{mo}}\le(f(a)-1\ri)\ri)}\label{eq:wt3}
\end{eqnarray}
where we have used $\rho_{mo}=\rho_{bo}+\rho_{\rm bdmo}$, we have again neglected DE, and $\Omega_{\rm xo}$ is today density parameter of the $x$ fluid. We see that $\omega_{\rm tot}$ is a function of $a$ and $x_c$ through $f(a,x_c)$ and $\omega_{\rm bdm}$ and this equation has to be solved numerically. We can rewrite Eq.(\ref{eq:wt3}) as,
\begin{equation}\label{eq:aeqbdm}
\fr{a}{a_{eq}}=\fr{(1-3\omega_{\rm tot})/3\omega_{\rm tot}}{1+\fr{\Omega_{bdmo}}{\Omega_{mo}}\le[f\le(1- \frac{\omega_{\rm bdm}}{\omega_{\rm tot}}\ri)-1\ri]}.
\end{equation}
Defining matter-radiation equality for BDM at $\tilde{a}_{eq}$ when $\omega_{\rm tot}=1/6$, which is also valid for the limiting case of the standard model, we notice that the quantity $(1-3\omega_{\rm tot}(a_{\rm eq}))/3\omega_{\rm tot}(a_{\rm eq}) = 1$. Also, it is interesting to note that $\tilde{a}_{\rm eq}> a_{eq}$ can be obtained without introducing extra relativistic particles and due to the time dependent EoS of BDM particles. Notice that in Eq.(\ref{eq:aeqbdm}), $\omega_{\rm bdm}$ and $f(a)$  depend  on $a_c$ and $v_c$  only through the combination $x_c=v_c a_c$. In the limit $x_c \rightarrow 0$ we have $\omega_{\rm bdm}=0, f=1$ and  $\tilde{a}_{eq}=a_{eq}$, as in standard CDM scenario with no extra degrees of freedom.

The connection between BDM and extra relativistic degrees of freedom at an arbitrary scale factor $a$  is easily achieve using Eq.(\ref{eq:aeq3}) and (\ref{eq:aeqbdm}), which yields
\begin{equation}\label{eq:nex}
\tilde{N}_{\rm ex}(a) =\frac{1+ \alpha N_\nu}{\alpha}\left( \frac{a}{a_{\rm eq}} \frac{3\omega_{\rm tot}(a)}{1-3\omega_{\rm tot}(a)} - 1 \right),
\end{equation}
and at equality we have $\omega_{\rm tot}=1/6 $ and $\tilde{N}_{\rm ex}(\tilde{a}_{eq}) =(\tilde{a}_{eq}/a_{\rm eq}  - 1 )(1+ \alpha N_\nu)/\alpha$.
The Eq.(\ref{eq:nex}) should be interpreted as giving an apparent number of extra relativistic particle $\tilde{N}_{\rm ex}$ at equality, even though we have not introduced extra particles, is due to the effect of a time dependent EoS for DM, i.e. for BDM. The $N_{\rm ex}$ without the ``tilde'' is for CDM scenario and it is constant by assumption. Because the BDM particles behaves as radiation before the epoch of the transition, $a<a_c$, the apparent number of extra relativistic neutrinos must remain constant. After the transition Eq.(\ref{eq:nex}) it is a function of the scale factor. This behavior is show in Fig.\ref{fig:Nexfdea}.

We can rearrange element of the last equation and extract the contribution of the BDM particles to the cosmic radiation $\rho_{\rm ex}(a) = \alpha N_{\rm ex}(a)\rho_\gamma(a)$ as a funtion of the scale factor
\begin{equation}\label{eq:rho_ex}
  \rho_{\rm ex}(a) = \frac{3\omega_{\rm tot}}{1-3\omega_{\rm tot}}\left( \rho_{\rm cdm} - \rho_{\rm bdm}\left[ 1 - \frac{\omega_{\rm bdm}}{\omega_{\rm tot}} \right] \right).
\end{equation}

We expect the phase transition to be $a_c < a_{eq}$, therefore our BDM would only be able to account for a small amount of $N_{ex}$. We plot in Fig.\ref{fig:omegabdm} the value of $\tilde{a}_{\rm eq}/a_{eq}$ and $N_{\rm ex}$ as a function of $x_c$. The largest amount of $N_{\rm ex}$ as a function of  $v_c$ is given at $v_c=1$, and since $x_c=v_ca_c$ if we set $v_c=1$ in Fig.\ref{fig:omegabdm} we have the upper level $x_a/a_{eq}|_{v_c=1}=a_c/a_{eq}$, and for example if $a_c/a_{eq}=1 $ we have a maximum amount of extra relativistic degrees of freedom $N_{\rm ex}=6.15$, for $a_c/a_{eq}=0.2 $ it reduces to $N_{\rm ex}=0.37$ while for $a_c/a_{eq}=0.1 $ we find $N_{\rm ex}=0.09$. The moment of the transition $x_c$ can be determined by CMB observations by the amount of matter, $\Omega_m$, and equality epoch, $z_{\rm eq}$ and in Fig.\ref{fig:CL} we show the 68\% and 95\% confidence level (C.L.) of $x_c$ considering Planck data.

\section{BBN}\label{sec:BBN}
\captionsetup{margin=5pt,font=small,labelfont=bf,justification=centerlast,indention=.25cm}
 \begin{figure}[t]
  \includegraphics[width=0.4\textwidth]{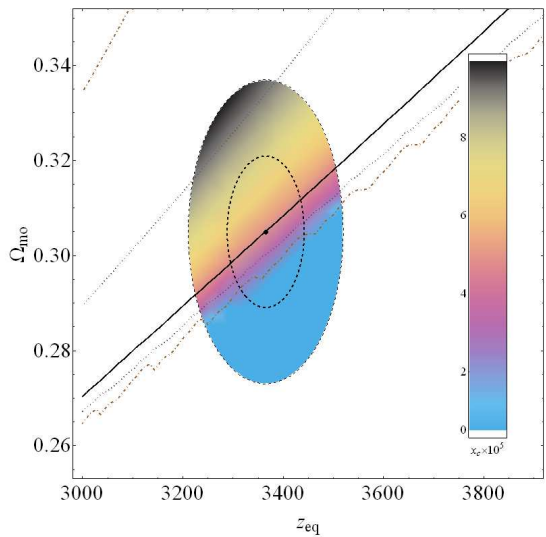}
  \caption{\footnotesize{ In this plot we show the degeneracy between $\Omega_{\rm mo}$ and $z_{\rm eq}$, in other words, different values of $\Omega_{\rm mo}$ and $z_{\rm eq}$ can give the same $x_c$. The black line represent $x_c = 4.13 \times 10^{-5}$ while the dotted green line represent the $\pm 1\sigma$ of $x_c$. The small (big) dotted circle represent the $1\sigma$($2\sigma$) C.L. between $z_{\rm eq}$ and $\Omega_{mo}$. The gradient color represent the different values that one gets with a different combination of $\Omega_{\rm mo}$ and $z_{\rm eq}$ ranging from $x_c > 10^{-8}$ (blue) to $x_c < 10^{-4}$ (black). The black dot represents the central value.}}
  \label{fig:CL2}
 \end{figure}

In the previous section we considered the observational results that extra radiation beyond the standard model is imprinted in the CMB  \cite{Hinshaw:2012fq}. Now, we explore the possibility that the BDM particles account for such an excess at BBN. The BDM particles can change the prediction of BBN for the abundance of the light elements, such as helium and deuterium by changing the radiation density at that epoch thereby increasing the expansion rate of the Universe during this stage of the Universe.

The $^4$He is very sensitive to the competition between the weak interaction rates and the expansion rate which, during the radiation dominated evolution is fixed by the energy density in relativistic particles. As a result $^4$He abundances tests the standard model and provides one of the strongest constraints on $x_c = a_c v_c$.

At the BBN epoch, before $e^{\pm}$ decoupling, the standard model of particle physics establish that energy density consists of an equilibrium mixture of photons, relativistic $e^{\pm}$ pairs, neutrinos, and antineutrinos as constituents of this dominant component. With all chemical potentials set to zero the energy densities are related by thermal equilibrium so that the total radiation density may be written in terms of the photon density as $\rho_{\rm r}^{\rm bbn} = \rho_\gamma + \rho_{e^\pm} + \rho_\nu = 43\rho_\gamma/8$, $\rho_{\rm ex}^{\rm bbn} =\fr{7}{8} N_{\rm ex}^{\rm \tiny bbn} \rho_\gamma $ and  ones has $T_\gamma=T_\nu$ at BBN.

It is convenient to defined the non-standard expansion rate as $S$, to account for the extra contribution to the standard-model energy density, as
\begin{eqnarray}\label{eq:s2_nex}
  S^2 = \left( \frac{\hat{H}}{H} \right)^2 = \left. \frac{\hat{\rho}_{\rm r}}{\rho_{\rm r}}\right\vert_{\rm \tiny bbn} = 1 + \fr{7}{43} N_{\rm ex}^{\rm \tiny bbn},
\end{eqnarray}
for the standard CDM considering extra radiation. This extra component is modeled just like an additional neutrino though we emphasize that the extra need may not be an additional flavors of active or sterile neutrinos it is just  additional relativistic degrees of freedom.

The following simple fits to the $^4$He mass fraction are quite accurate and takes into account the nonstandard expansion \cite{Steigman:2008eb,Komatsu:2010fb},
\begin{equation}\label{eq:bbn_yp}
  Y_p = 0.2485 + 0.0016 \left[ \left(\eta_{10} - 6\right) + 100 \left( S - 1 \right) \right]
\end{equation}
where $\eta_{10}= 273.9 \Omega_b h^2$ is the baryon to photon ratio. The last equation is the connection between the neutral hydrogen, the BDM model, and the extra relativistic degrees of freedom at the time of BBN, $N_{\rm ex}^{\rm bbn}$. If $S \neq 1$ ($N_{\rm ex}^{\rm bbn} > 0$) it is an indication of new physics beyond the standard model.

Using the values of $\Omega_m$, $z_{\rm eq}$ we can determine the value of $N_{\rm ex}(z_{\rm eq})$ at equality (c.f. Eq.(\ref{eq:aeq3})) and with $Y_p$ we constrain $N_{\rm ex}^{\rm \tiny bbn}$ at BBN (c.f. Eq.(\ref{eq:s2_nex})). From the Table \ref{tab:data} we can see that the central values for Planck$_\star$, i.e. $\Omega_m=0.305, z_{\rm eq}=3365, Y_p=0.26$,  gives $N_{\rm ex}(z_{\rm eq})= 0.14$ at equality and  $N_{\rm ex}^{\rm \tiny bbn}= 0.90$ at BBN. Clearly the values of $N_{\rm ex}$ at equality and BBN are quite different and in a CDM scenario one should have a constant $N_{\rm ex}$, i.e. $N_{\rm ex}^{\rm \tiny bbn}= N_{\rm ex}(z_{\rm eq})= N_{\rm ex}(a_{o})$, if the particles are still relativistic at equality and/or at present time. However, the value of  $N_{\rm ex}^{\rm \tiny bbn}$ is very sensitive to $Y_p$ and   $N_{\rm ex}^{\rm \tiny bbn}=N_{\rm ex}(z_{\rm eq})=0.14$ requires an $Y_p=0.2505$.

Let us now study the constrains on BDM from BBN. Since we expect that the phase transition of BDM takes place after BBN, the BDM particles are relativistic during the nucleosynthesis epoch. Using  Eqs.(\ref{eq:rho_bdm}) and  (\ref{rc}) we have for $a<a_c$
\begin{equation}\la{rhoc}
    \rho_{\rm bdm} = \rho_{\rm c} \left( \frac{a}{a_c} \right)^{-4} \simeq \rho_{\rm bdmo} \frac{a_c}{a_o}\left( \frac{a}{a_o} \right)^{-4} e^{v_c^2/2},
\end{equation}
Therefore, the non-standard expansion rate become
\begin{equation}\label{eq:bbn_s2}
  S^2 = \left( \frac{\tilde{H}}{H} \right)^2 = \frac{\tilde{\rho_r}}{\rho_r} = 1 + \frac{8}{43} \frac{\rho_{\rm bdm}}{\rho_\gamma}
\end{equation}
From Eqs.(\ref{eq:s2_nex}), (\ref{rhoc}) and (\ref{eq:bbn_s2}) and using the fact that at BBN $\rho_\gamma=  \rho_{\gamma o} (T_{\rm \gamma,bbn}/T_{\gamma o})^4$  with $T_{\rm\gamma,bbn}/ T_{\gamma o} = (a_o/a)(g_o/g_{\rm \tiny bbn})^{1/3} $ and $g_o/g_{\rm \tiny bbn}=4/11$ is the ratio of the degrees of freedom of the relativistic components in thermal equilibrium with the photons at present time ($g_o=2$) and   just after neutrino decoupling ($g_{\rm \tiny bbn}=11/2$) and before $e^+  e^-$ annihilation . Therefore we have
\begin{equation}\label{eq:nex_bdm}
 N_{\rm ex}^{\rm bbn}= \frac{8}{7}\frac{\rho_{\rm bdm}}{\rho_\gamma} = \frac{1+\alpha N_\nu}{\alpha}\frac{\Omega_{bdmo}}{\Omega_{m o}}\frac{a_c}{a_{eq}}\,e^{v_c^2/2}
\end{equation}
where we have also used $\Omega_{m o}/\Omega_{r o}=a_o/a_{eq}$ and $\rho_{r o}=(1+\alpha N_\nu)\rho_{\gamma o}$.
The BDM particles are relativistic above $E_c$, i.e for $a<a_c$, the number of  $N_{\rm ex}(a\leq a_c)=N_{\rm ex}(a_c)$ remains constant
as seen in Fig.\ref{fig:Nexfdea}. This includes the time of BBN, so BDM must have  $N_{\rm ex}^{\rm bbn}=N_{\rm ex}(a_c)$.

\section{Results}\label{sec:analysis}
\captionsetup{margin=5pt,font=small,labelfont=bf,justification=centerlast,indention=.25cm}
\begin{figure}[t]
  \includegraphics[width=0.45\textwidth]{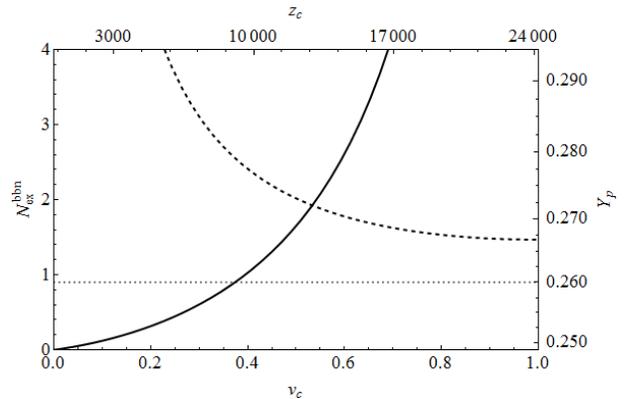}
  \caption{\footnotesize{Plot of the extra relativistic degrees of freedom as a function of $v_c$. The thick line is $\tilde{N}_{\rm ex}(a_c)$ evaluated at the time of the transition, Eq.(\ref{eq:nex}). The dashed line is $N_{\rm ex}^{\rm bbn}$, Eq.(\ref{eq:nex_bdm}). The dotted line is the extra relativistic degree of freedom correspondent to the reported value of $Y_p$, c.f. Eq.(\ref{eq:s2_nex}). The first line is derived from the extra radiation at the time of equality making $\omega_{\rm tot} = 1/6$. The second line is from the constrains of the nonstandard expansion rate at the BBN epoch. In all cases we assume a fixed value for $x_c = 4.13 \times 10^{-5}$. The BDM model predict a value for $Y_p = 0.272$ which is within the $1\sigma$ error of the reported $Y_p = 0.26$, see the discussion in Sec.\ref{sec:BBN} and \ref{sec:Discussion} for more details.}}
  \label{fig:NexVsVc}
\end{figure}

In this section we compute the values of the BDM parameters $x_c$ and $v_c$ form current cosmological data. We use the
published values of $\neff$ to determine   $x_c$ and $v_c$ at equality and BBN. From Eqs.(\ref{eq:aeqbdm}) and (\ref{eq:nex}), with $a_{eq}\ll a_o$,  we see that $\rho_{\rm bdmo}$ and  $N_{\rm ex}$  are determined only by the value of $x_c=a_c v_c$ and the amount of DM at present time $\Omega_{\rm bdmo}$. On the other hand, at BBN the amount of $Y_p$ given in Eqs.(\ref{eq:nex_bdm})  depends on $\Omega_{\rm bdmo}, x_c$ and also on $v_c$. Therefore, we can  constrain $x_c$ and $v_c$ using the value of $\neff$ at these different epochs.

We see from Table \ref{tab:data} that WMAP9, ACT and SPT results have  at $1\sigma$  an $N_{\rm ex} > 0$ but Planck data has an $N_{ex}=0$ at  $1\sigma$. However, the central value of $N_{ex}$ hints for a small amount of extra relativistic degrees of freedom and  its value is highly dependent on $H_o$ \cite{Ade:2013lta}. The larger $H_o$ the more amount of relativistic degrees of freedom is required and therefore we present our results using different sets of data.
An $N_{ex}>0$ implies the need of extra relativistic particles for a CDM cosmology or a non vanishing  value of  $x_c$ in BDM. A small  value of $N_{\rm ex}$ requires a very small $x_c$ and for $N_{\rm ex} \leq 0.07$ then $x_c$ will be constrained to be less than $x_c < 2 \times10^{-5}$ (for $v_c=1$ we get $a_c/a_{eq}<0.09$)  and  if $\neff \simeq  N_\nu$ then $x_c \ll 10^{-5}$ as can be seen in Fig.\ref{fig:omegabdm}. From Eq.(\ref{eq:aeqbdm}) and Fig.\ref{fig:CL2} we see that we can get the same $x_c$ with a combination of different values of $\Omega_m$ and $a_{\rm eq}$. Hence, we use the results with the strongest constraints on $\Omega_m$, namely, we use results from the combined data analysis of CMB, BAO and $H_o$ when available, c.f. Table \ref{tab:data}.

Using the relation between $N_{\rm ex}$ and $x_c$ in Eq.(\ref{eq:nex}) at equality epoch  and ${\rm Planck}_\star$ results, the cosmological observations give a value
\begin{equation}\label{eq:result_zeq}
  x_c = 4.13^{+3.65}_{-4.13} \times 10^{-5} .
\end{equation}
% Figure of CL - aqui hablar de todas las fig en aeq 1 y CL
Fig.\ref{fig:CL} shows the 68\% and 95\% C.L. of $x_c$ considering Planck data. The moment of the transition $x_c$ can be determined by CMB observations by the amount of matter, $\Omega_m$, and equality epoch, $z_{\rm eq}$. The contours lie on the expected linear correlation between $\Omega_{\rm m}$ and $a_c$ given by Eq. (\ref{eq:omega_total}) for which we take the value show in Table \ref{tab:data}.

We now constrain from BBN  the value of $x_c=a_c v_c$ and $v_c$ using Eq.(\ref{eq:nex_bdm}) and (\ref{eq:bbn_yp})  giving a value
\begin{eqnarray}\label{eq:bbn_xc}
   \frac{x_c}{v_c}\,e^{v_c^2/2} = 4.17^{+6.86}_{-4.17} \times 10^{-5} .
\end{eqnarray}
If we take  the  previous result on $x_c$  (Eq.(\ref{eq:result_zeq})) at equality we can determine the value of $v_c$. Notice that the dependence on $v_c$ in Eq.(\ref{eq:bbn_xc}) is given by the quantity $g(v_c)\equiv e^{v_c^2/2}/v_c$   which has a lower limit $e^{1/2}=1.64\leq g(v_c=1)$ since the velocity must be between $0\leq v_c\leq 1$. Therefore,  the central value of Eq.(\ref{eq:bbn_xc}) gives an upper value to $x_c < 2.7 \times 10^{-4}$ for $v_c=1$ which is an order of magnitude larger than $x_c$ in Eq.(\ref{eq:result_zeq}).  However, we expect to have $v_c<1$ if the BDM mass is due to non-perturbative physics as suggested for BDM.

Consistency of BDM requires that the apparent number of extra degrees of freedom at BBN is the same as at the time of the BDM at the transition $a_c$, since BDM particles are relativistic for $a\leq a_c$. At a fixed value of $x_c$ we have
that $\tilde{N}_{\rm ex}(a_c)$ and $N_{\rm ex}^{\rm bbn}$, in Eqs.(\ref{eq:nex}) and (\ref{eq:nex_bdm}), are only a function of $v_c$.  In Fig.\ref{fig:NexVsVc} we plot the value of $\tilde{N}_{\rm ex}(a_c)$ and $N_{\rm ex}^{\rm bbn}$ as a function of $v_c$ and  the result for $v_c$ at $\tilde{N}_{\rm ex}(a_c)=N_{\rm ex}^{\rm bbn}$ is shown in Table \ref{tab:results} for the different data.
The extra radiation due to the BDM particles change the amount of neutral hydrogen produced at BBN epoch. 

The value of $Y_p$ can be predicted for the BDM model assuming that the apparent number of extra degrees of freedom evolves as in Eq.(\ref{eq:nex}). Replacing $\tilde{N}_{\rm ex}(a_c)$ in Eq.(\ref{eq:nex_bdm}) we are able to constrain the value of $v_c$ and $Y_p$ simultaneously knowing only the moment of equality. Therefore, BDM  relates  the amount of neutral hydrogen produced at BBN with  equality epoch, and viceversa. In this case, taking Planck$_\star$ data ($z_{\rm eq} = 3365$)  gives an $N_{\rm ex}^{\rm bbn} = 1.92$ and $v_c = 0.53$ and BDM requires an $Y_p = 0.272$ (within $1\sigma$ C.L.), see Fig. \ref{fig:NexVsVc}. We would like to emphasis that the value of $Y_p $ and  $N_{\rm ex}^{\rm bbn}$ is quite sensitive to $x_c$ and for example if we take $x_c=10^{-5}$ then we get   $Y_p=0.256 $ and  $N_{\rm ex}^{\rm bbn}=0.62$.

\begin{figure}[b]
    \subfloat[\footnotesize{WMAP9}]{ \includegraphics[width=0.24\textwidth]{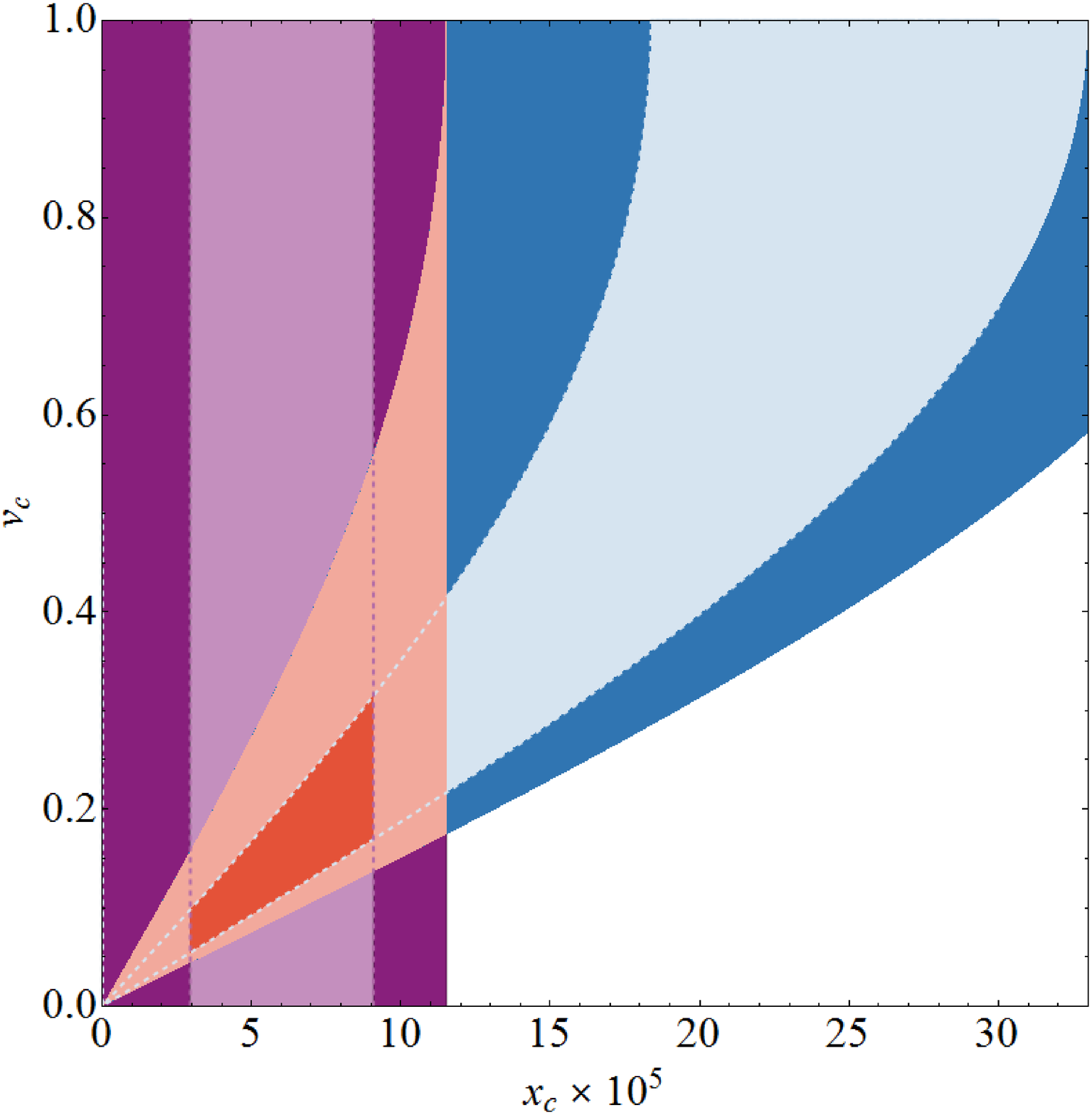}}
    \subfloat[\footnotesize{Planck$_\star$}]{\includegraphics[width=0.24\textwidth]{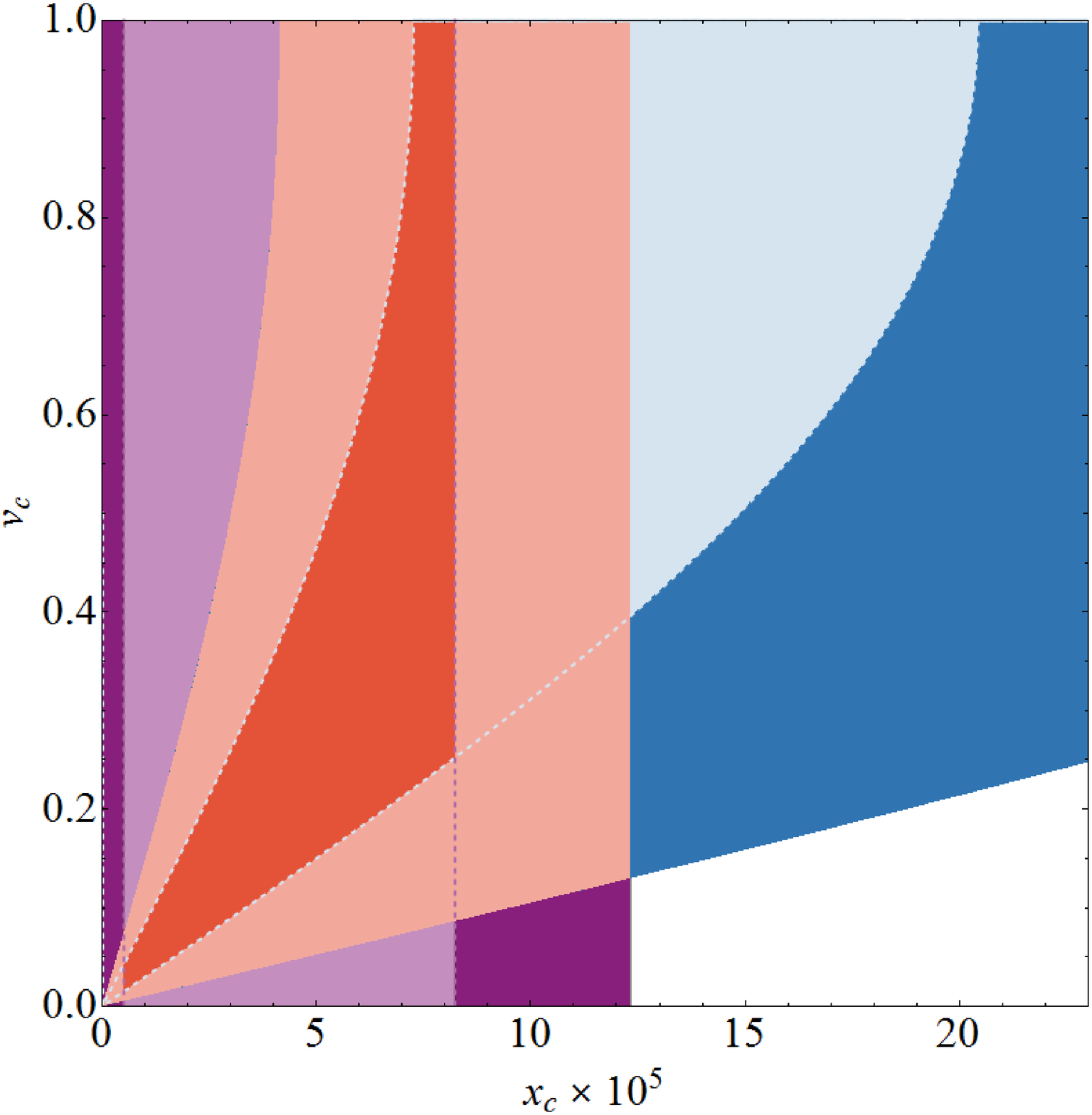}}
  \caption{\footnotesize{ In this plot we show the range of values valid for $x_c$ and $v_c$ given the constraints of $^4$He and $z_{\rm eq}$ and using WMAP9 (left panel) and ${\rm Planck}_\star$ (right panel) data (c.f. Table \ref{tab:results}). The purple and blue regions represent the C.L. (68\% and 96\%) obtained with $\neff$ (c.f. Sec.\ref{sec:analysis}) and $^4$He data, respectively. The orange region is where the result of both analysis overlaps. Notice that we cannot constrain the value of $v_c$ using only the $\neff$ data.   }}
  \label{fig:CL4}
\end{figure}

In Table \ref{tab:results} we summarize the constraints on the BDM parameters $x_c$ and $v_c$ obtained directly from the $\neff$ and $^4$He using different previous results, such as ${\rm Planck}_\star$, WMAP9, ACT, and SPT. As well as the derived parameters such as the moment, $1+z_c = a_c^{-1}$, the energy of the BDM particles, $E_c=\rho_c^{1/4}$ (c.f. Eq.(\ref{rc})), and the energy of the Universe $E\equiv \rho^{1/4}$, with $\rho$ the total energy density, the last two quantities at the moment of the transition.  Notice that with Eq.(\ref{eq:bbn_xc}) and the constrain on $x_c$ (c.f. Eq.(\ref{eq:result_zeq})) we are able to derive the constraints of the central value of  $x_c = 4.13 \times 10^{-5}$, $v_c = 0.37$. However, at $1\sigma$ Planck data have that $N_{ex}=0$ therefore $x_c = 0$ is valid a $1\sigma$. Hence, the moment of the transition would only be constraint to  $z_c \geq 6445$ and the energy of the Universe at $E(a_c) \geq 2.65$ {\rm eV}. In Fig.\ref{fig:CL4} we show the range of values at $1\sigma$ and $2\sigma$ (68\% and 95\% C.L. region) valid for $x_c$ and $v_c$ combining the two evidence for extra radiation, the one steaming from equality epoch and  the  amount of primordial $^4$He.

%there would be no DM to account for structure formation
\captionsetup{margin=5pt,font=small,labelfont=bf,justification=centerlast,indention=.25cm}
  \setlength{\tabcolsep}{0.05cm}
  \def\arraystretch{1.25}
  \ctable[
     caption = {We present the constraints on $x_c$ and $v_c$ as discussed in Sec. \ref{sec:framework} and \ref{sec:BBN} using different result on $\neff$ and $Y_p$ (c.f.  Table \ref{tab:data}). We also present the moment ($z_c$) and the energy when the transition happens ($E_c$). Notice that in some cases the transition $z_c < z_{\rm eq}$, however we expect these cases not to be valid in order to account for structure formation. We present the minimum value for $z_c$ and $E_c$ for the ${\rm Planck}_\star$ data because no extra radiation is contained at $1\sigma$ therefore the moment of the transition should be consistent with $x_c \rightarrow 0$ at $1\sigma$.},
     label=tab:results,
    captionskip=\smallskipamount]{lccccc}
    {  \tnote[a]{This results considered Planck+WP+highL+BAO combined data. \cite{Ade:2013lta}.} }
    {
                                   &${\rm Planck}_\star$\tmark[a] & \footnotesize{WMAP9} & \footnotesize{ACT} & \footnotesize{SPT} \\
      \hline
      \hline
      $x_c \times 10^5$  .       & $4.13^{+3.65}_{-4.13}$ & $6.64^{+2.46}_{-3.54}$       & $6.34^{+2.56}_{-4.16}$       & $8.25^{+2.92}_{-4.07}$ \\
      $N_{\rm ex}^{\rm bbn}$ . . . & $0.9^{+1.5}_{-0.9}$    & $2.4^{+1.6}_{-1.4}$          & $0.6^{+1.8}_{-0.6}$          & $4.2^{+1.6}_{-1.5}$ \\
      $v_c$  . . . . .           & $0.37^{+0.18}_{-0.17}$ & $0.54^{+0.09}_{-0.10}$       & $0.26^{+0.13}_{-0.14}$       & $0.64^{+0.09}_{-0.10}$  \\
      $z_c$  . . . . .           & $\geq 24217$  & $15060^{+2\times10^4}_{-1\times10^4}$ &$15781^{+3\times10^4}_{-6\times10^3}$ &$6313^{+10^4}_{-3\times10^3} $\\
      $E_c [eV]$  . .            & $\geq 3.89 $           & $2.75^{+2.12}_{-1.77} $      & $2.38^{+3.49}_{-1.03}  $     & $1.32^{+2.60}_{-0.55} $\\
      $E(a_c)[eV]$               & $\geq 9.01 $           & $5.93^{+5.77}_{-3.9}$        & $6.17^{+9.87}_{-2.34} $      & $2.65^{+6.32}_{-1.13}$ \\
      \hline
  }

\section{Conclusion} \label{sec:Discussion}
\captionsetup{margin=5pt,font=small,labelfont=bf,justification=centerlast,indention=.25cm}

Cosmological observations suggest the existence of extra radiation, $\neff > N_\nu$, in order to explain CMB and $^4$He measurements. Motivated by this lack of radiation in the standard CDM framework we have considered the BDM model, which may explain the need for an extra relativistic component without introducing new particles.

The BDM particles behave as radiation for scales $a < a_c$, while for  $a > a_c$ these particles become non-relativistic due to a phase transition in which the particles acquire a mass due to non-perturbative methods, when $v_c < 1$, as protons and neutrons do. We expect this phase transition to be between BBN (E=$\mathcal{O}$(MeV)) and matter radiation equality (E=$\mathcal{O}$(eV)). If $E\gg MeV$ then our BDM will be indistinguishable from CDM and for $E<eV$ BDM would not be able to account for structure formation. The evolution of the BDM energy density  during this process is described by a time dependent EoS, $\omega_{\rm bdm}(a)$. The amount of radiation due to the transition of the BDM particles change as function of the scale factor $\rho_{\rm ex}(a)$, c.f. Eq.(\ref{eq:rho_ex}), if this extra radiation is modeled as neutrino radiation this means that the neutrino relativistic degrees of freedom will be also a function of the scale factor $N_{\rm ex}(a)$.

Since we have a time dependent EoS then we cannot simply use $\rho_r = \rho_{m}$ to determine matter-radiation equality.  Instead we define equality when the total EoS is $\omega_{\rm tot} = 1/6 $ which overcomes the fact that $\omega_{\rm bdm}$ is a function of $a$ and is also valid in the limiting case of the standard model. We conclude that the apparent number of relativistic particle, $N_{\rm ex}$, is explained by a time dependent EoS of the DM without introducing new particles, c.f. Sec.\ref{sec:framework}. For a phase transition $a_c\ll a_{eq}$ the amount of apparent extra relativistic degrees of freedom in our BDM model is small and for $N_{ex}\leq 0.07$ one requires $a_c/a_{eq}\leq 0.09$ if $v_c=1$. The BDM particles also change the prediction of BBN for the abundance of the light elements, such as helium, by changing the radiation density thereby increasing the expansion rate of the early Universe, incidentally, observation also shows an excess during BBN which can be explained by the Dark Matter BDM  particles.

We compute the range of values for the transition epoch $x_c = a_c v_c$  and $v_c$ using cosmological data  which predict extra radiation, Table \ref{tab:results} summarize the results. Using the latest result of Planck$_\star$, we conclude that the order of the transition should be $x_c = 4.13(^{+3.65}_{-4.13}) \times 10^{-5}$ in order to explain the evidence of extra radiation at matter-radiation equality $a_{\rm eq}$. Using $^4$He results of the BBN we obtain equivalent constrain for $N_{\rm ex}^{\rm bbn} = 0.9^{+1.5}_{-0.9}$. Combining both previous results we are able to constrain the velocity $v_c = 0.37^{+0.18}_{-0.17}$, and therefore $z_c > 24217$ and an $E(a_c)\ge9.01$. However, if the value of $\neff$  becomes close to $N_\nu$, i.e. $\neff \simeq  N_\nu$, then $x_c \ll 10^{-5}$ and  $z_c\gg 10^{5}$.

The BDM model also is able to explained the inconsistency between the apparent extra degrees of freedom at equality and BBN epoch, $N_{\rm ex}(a_{\rm eq}) \neq N_{\rm ex}^{\rm bbn}$, and predict the amount of $^4$He given the moment of equality $z_{\rm eq}$, and viceversa. From the assumption that equality occurs when $\omega_{\rm tot} = 1/6$ we where able to compute how $\tilde{N}_{ex}$ is dependent of the scale factor, Eq.(\ref{eq:nex}). Combining this equation with the one obtained from BBN $N_{\rm ex}^{\rm bbn}$, Eq.(\ref{eq:nex_bdm}), we where able to predict that the amount of $^4$He consistent with a $z_{\rm eq} = 3365$ and  $x_c=4.13\times 10^{-5}$  should be $Y_p = 0.272$, which is conciliable within the $1\sigma$ error of reported $Y_p = 0.26$ but a slightly smaller $x_c=10^{-5}$ gives  $N_{\rm ex}^{\rm bbn}=0.62$ and $Y_p=0.256 $.

We conclude that we can account for the apparent extra $N_{ex}$ at equality and BBN epochs using only the BDM particles which have a time dependent EoS $\omega_{\rm bdm}(a)$,  with no need to introduce extra relativistic particles. However, further analysis will provide us with a better understanding of dark matter and the possibility that the dark matter mass is due to non perturbative physics.

\section{Acknowledgment}
We acknowledge financial support from Conacyt Project No. 80519 and UNAM PAPIIT Project No. IN100612. We thanks to Prof. J.Cervantes-Cota for his helpful comments and discussions.

\bibliographystyle{apsrev4-1}
\bibliography{ExtraRad}

\end{document}